\documentstyle[preprint,psfig,aps]{revtex}

\begin{document}
\baselineskip=20pt
\def\gtsim{\vbox {\hbox{\lower 0.9\baselineskip \hbox{$>$}} \break
                 \hbox{\lower 0.2\baselineskip \hbox{$\sim$}} } }

\title{On Magnetic Impurities in Gapless Fermi Systems}
\author{L.S. Borkowski}
\address{Institute of Physics, A. Mickiewicz University,\\
Matejki 48/49, 60-769 Pozna\'n, Poland}

\maketitle

\vskip 0.3cm

\begin{abstract}
In ordinary metals, antiferromagnetic exchange between conduction
electrons and a magnetic impurity leads to screening of the impurity
spin below the Kondo temperature, $T_K$. In systems such as semimetals,
small-gap semiconductors and unconventional superconductors, a reduction
in available conduction states near the chemical potential can greatly
depress $T_K$. The behavior of an Anderson impurity
in a model with a power-law density of states,
$N(\epsilon) \sim |\epsilon|^r$, $r>0$, for $|\epsilon| < \Delta$,
where $\Delta \ll D$, is studied using the non-crossing approximation.
The transition from the Kondo singlet to the magnetic ground state
can be seen in the behavior of the impurity magnetic susceptibility $\chi$.
The product $T\chi$ saturates at a finite value at low temperature
for coupling smaller than the critical one. For sufficiently
large coupling $T\chi \rightarrow 0$, as $T \rightarrow 0$,
indicating complete screening of the impurity spin.
\end{abstract}

\vskip 0.5cm

\newpage

\noindent
{\it Introduction}.
In a number of Fermi systems the density of states $N(\epsilon)$
vanishes at the Fermi surface $E_F$ and varies linearly or
quadratically for $|\epsilon|/D \equiv |E-E_F|/D \ll 1$,
where $D$ is the energy scale associated with the conduction
electron bandwidth. This situation may arise e.g. in
heavy-fermion or cuprate superconductors and anisotropic
heavy fermion semiconductors.\cite{Miyake1}
Also exotic phases of the Hubbard model may possess
$N(\epsilon) \sim |\epsilon| $ in two dimensions.\cite{AffMar}

In normal metals dilute impurities coupled antiferromagnetically
to the conduction band lead to low-temperature reduction
of the Curie term in the impurity magnetic susceptibility and
an increase in the resistivity. This is known as a Kondo effect.
The formation of the spin-singlet state favored by
the antiferromagnetic coupling depends on the availability
of electronic states at low energies.

Earlier studies by poor-man's scaling and large-$N$ method,
where $N$ is impurity orbital degeneracy,
showed that the Kondo effect survives if the coupling
between electrons and the impurity $J$ is larger than
a critical value $J_c$.\cite{Fradkin}
In a gapless system with $N(\epsilon) \sim |\epsilon|^r$,
$J_c$ scales linearly with $r$ for $r \ll 1$.
A large-$N$ approach to magnetic impurities
in superconductors\cite{lsb1,lsb2} leads to similar results for $J_c$.
However, for $r \leq 1$ or $N=2$, any finite impurity
concentration was found to result in $J_c=0$.
Numerical renormalization group calculations \cite{Kevin2,Chen}
and third-order scaling\cite{Kevin1} show that the Kondo effect
does not occur for $r > 1/2$
in the particle-hole symmetric
problem. Breaking this symmetry e.g. by potential
scattering or band asymmetry helps the screening
of the impurity moment. The critical coupling $J_c$ was found to be
strongly dependent on the magnitude of the potential scattering
term.\cite{Kevin2}
Earlier calculations for the case of a full gap, $N(\epsilon)=0$ for
$|\epsilon| < \Delta \ll D$, also found finite $J_c$ away from particle-hole
symmetry.\cite{Takegahara,Satori}

In this work the SU($N$) Anderson model
is studied in the non-crossing approximation
(NCA). In the limit of large Coulomb repulsion $U$
on the impurity site and for temperatures $T \ll U$,
the model has the form,

\begin{eqnarray}
H & = & \sum_{k,m} \epsilon_k c_{km}^\dagger c_{km}
+E_f \sum_{m} f_m^\dagger f_m
  + V\sum_{k,m}(c_{km}^\dagger f_m b^+ + h.c.)
  +  \lambda (\sum_m f_m^\dagger f_m + b^+b -1) ~,
\end{eqnarray}
where $E_f$ is the position of the bare impurity level, $f$ and $b$
are the impurity fermion and the slave boson operators, respectively.
The last term in the Hamiltonian follows
from the restriction of the Hilbert space to a singly occupied
impurity site, $\sum_m f^\dagger_m f_m + b^+b = 1$, $m=1,...,N$.
The self-energies of the slave boson and the impurity fermion
Green's functions are given by

\begin{equation}
\Sigma_0(\omega+i0^+)=NV^2 \int_{- \infty}^{\infty} d\epsilon
f(\epsilon) N(\epsilon) G_m(\omega+\epsilon+i0^+) ~,
\end{equation}

and

\begin{equation}
\Sigma_m (\omega+i0^+) = V^2 \int_{- \infty}^{\infty} d\epsilon
(1-f(\epsilon)) N(\epsilon) G_0(\omega-\epsilon+i0^+) ~.
\end{equation}

The density of states of the conduction band is assumed to be of
the form $N(\epsilon) = C|\epsilon/\Delta|^r \exp(-(\epsilon/D)^2)$ for
$0 < |\epsilon| < \Delta/D$, and $C \exp(-(\epsilon/D)^2)$
otherwise, and $C$ is a normalization constant.
The exponential part of $N(\epsilon)$ does not influence
the low-energy physics in any important way, while
it is convenient in solving the integral equations (2) and (3).

\noindent
{\it Numerical results}.
Here we focus on the non-degenerate case,
$N=2$. Results for static spin susceptibility are shown
in Figure 1 for $\Delta/D=10^{-5}$, $E_f/D=-0.67$, and $r=1$ and $r=2$.
For larger $\Gamma \equiv \pi N_0 V^2$, $T\chi$ decreases to zero
at low temperature, which is associated with the screening
of the impurity spin. For $\Gamma$ smaller than a certain critical
coupling $\Gamma_c$, $T\chi$ remains finite, as $T \rightarrow 0$,
indicating that impurity is not screened.
The critical coupling for the data sets presented in Fig. 1
is $\Gamma_c/D \simeq 0.108$ for $r=1$ and
$\Gamma_c/D \simeq 0.115$ for $r=2$.
Qualitatively similar behavior of the impurity susceptibility
was found by numerical renormalization
group calculations.\cite{Kevin2,Chen}

The transition from the spin-singlet ground state to unscreened
moment is also reflected in the impurity density of states.
The Abrikosov-Suhl resonance approaches the Fermi level when
$\Gamma \rightarrow \Gamma_c$. For $\Gamma < \Gamma_c$
the resonance falls below $E_F$ as illustrated in Figure 2.
Analogous behavior of $N_f(\omega)$ was noted earlier by Ogura
and Saso\cite{Saso} for the case of a full gap ($r=\infty$).

Preliminary analysis of the dependence of the critical coupling
on $\Delta$ in the limit $\Gamma \ll -E_f$, and $\Delta/D \ll 1$,
indicates scaling $\Gamma_c \sim D/\ln(D/\Delta)$,
independent of $r$, at least for $r \ge 1$.
This can be expected on the basis of the large-$N$ mean-field
results in the Kondo limit,\cite{lsb1} where it was found
that the critical exchange coupling is
$J_c \simeq 2D/\ln(2D/\Delta)$ for $\Delta \ll D$
in a model with $N(\epsilon)=const$ outside the pseudogap
region.

A more detailed study, including results
for $N > 2$, will be presented in a separate publication.

{\it Acknowledgements.} I would like to thank
K. Ingersent, P.J. Hirschfeld, A. Schiller and M. Hettler
for discussions. This work was supported in part by
KBN (Poland) grant no. 2P03B05709 and
by ISI Foundation and EU PECO Network ERBCIPDCT940027.
Numerical work was perfomed at the Pozna\'n Supercomputer
Center.

\newpage

\begin{figure}
\caption{Impurity spin susceptibility $T\chi$ as a function
of $\log(T/D)$ for $r=1$ and $r=2$. The magnitude of the pseudogap
is $\Delta/D = 10^{-5}$ and the bare impurity level is $E_f/D=-0.67$.}
\end{figure}

\begin{figure}
\caption{The low-energy part of the impurity density of states
$N_f(\omega)$ for $r=1$ and the same data set as in Figure 1,
evaluated at $T/D=2\times 10^{-7}, 1.2\times 10^{-7},
1.4\times 10^{-7}$, and $2\times 10^{-7}$ for
$\Gamma/D=0.10, 0.105, 0.11$, and $0.12$, respectively.}
\end{figure}

\vfil\eject
{}~

{}~

{}~

{}~

{}~

{}~
\begin{figure}[htp]
\leavevmode\centering\psfig{file=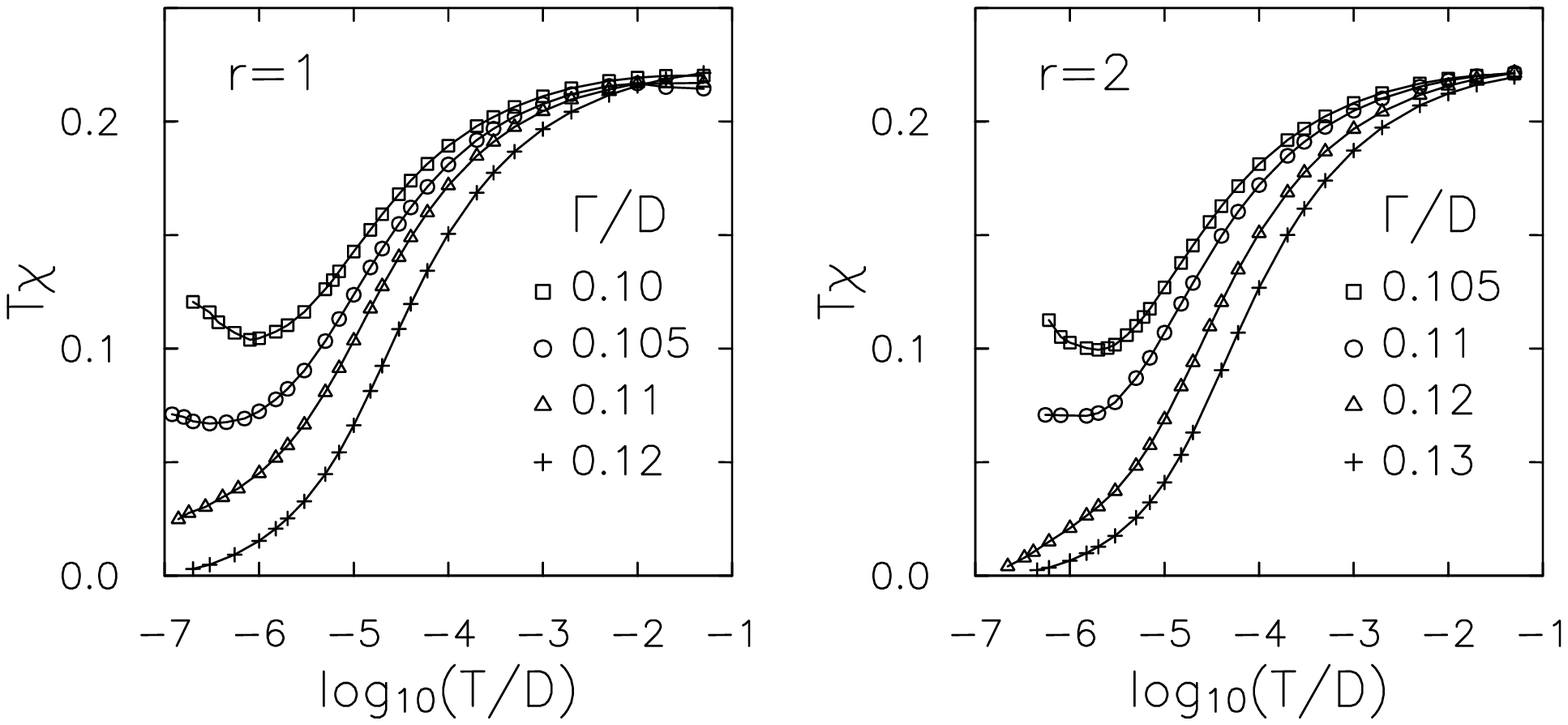,height=7.cm}
\end{figure}

\vfil\eject
{}~

{}~

{}~

{}~

{}~

{}~
\begin{figure}[htp]
\leavevmode\centering\psfig{file=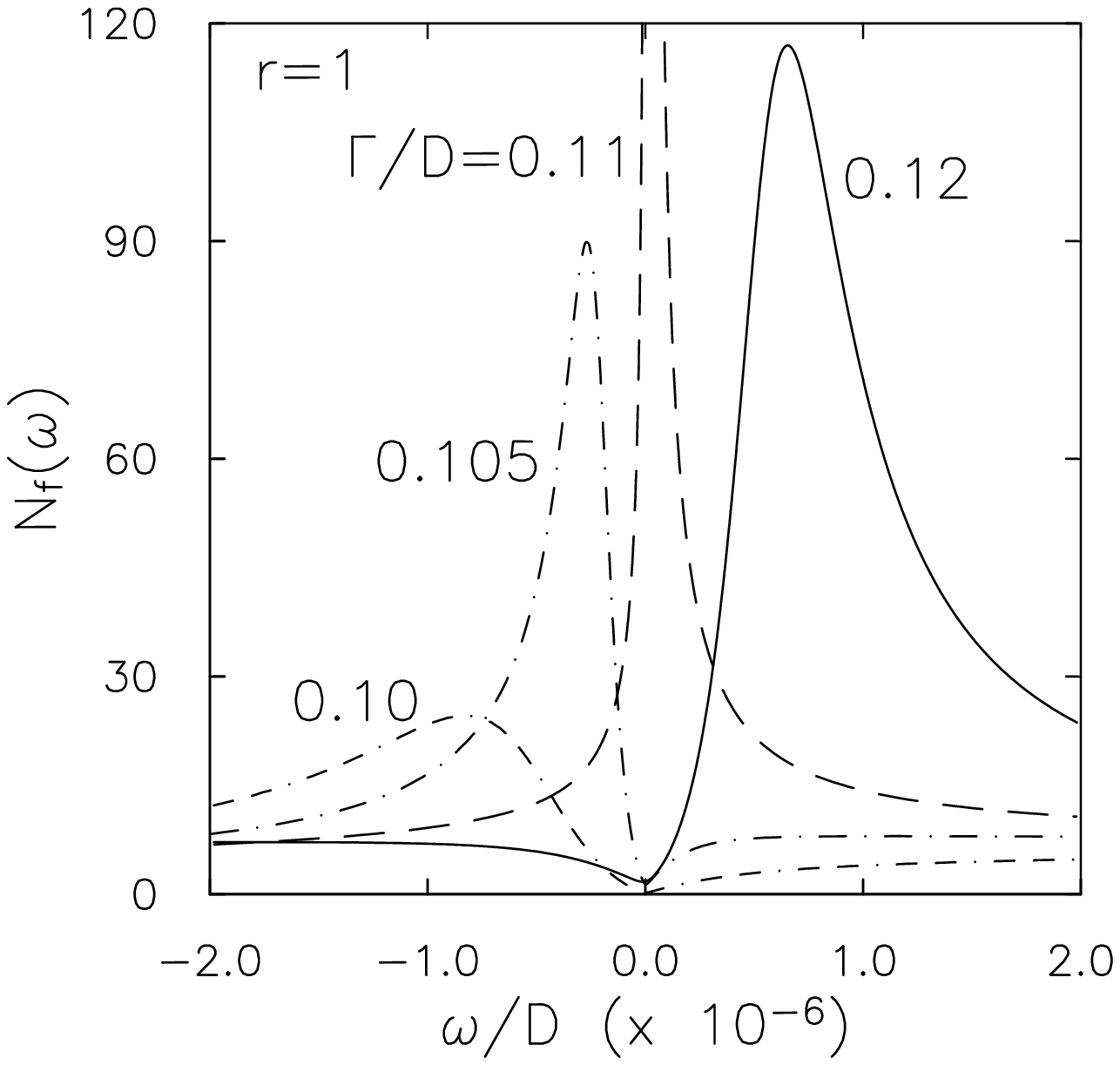,height=10.cm}
\end{figure}

\end{document}